\documentclass[sigconf]{acmart}
\AtBeginDocument{%
  \providecommand\BibTeX{{%
    \normalfont B\kern-0.5em{\scshape i\kern-0.25em b}\kern-0.8em\TeX}}}

\usepackage{hyperref} 

\copyrightyear{2022}
\acmYear{2022}
\setcopyright{acmcopyright}\acmConference[IWSiB'22 ]{5th International Workshop on Software-intensive Business: Towards Sustainable Software Business}{May 18, 2022}{Pittsburgh, PA, USA}
\acmBooktitle{5th International Workshop on Software-intensive Business: Towards Sustainable Software Business (IWSiB'22 ), May 18, 2022, Pittsburgh, PA, USA}
\acmPrice{15.00}
\acmDOI{10.1145/3524614.3528632}
\acmISBN{978-1-4503-9302-7/22/05}

\begin{document}

\title{Towards Understanding Analytics in Software Startups}
\author{Usman Rafiq}
\orcid{0000-0003-3198-851X}
\affiliation{%
  \institution{Faculty of Computer Science, Free University of Bozen-Bolzano, }
  \city{Bolzano}
  \country{Italy}
  \postcode{39100}
}
\email{urafiq@unibz.it}

\renewcommand{\shortauthors}{Usman Rafiq}

\begin{abstract}
Analytics plays a crucial role in the data-informed decision-making
processes of modern businesses. Unlike established software companies, software startups are not seen utilizing the potential of
analytics even though a startup process should be primarily data-driven. There has been little understanding in the literature about
analytics for software startups. This study set out to address the
knowledge gap by exploring how analytics is understood in the context of software startups. To this end, we collected the qualitative
data of three analytics platforms that are mostly used by startups
from multiple sources. We covered platform documentation as well
as experience reports of the software startups using these platforms. The data was analyzed using content analysis techniques.
Four high-level concepts were identified that encapsulate the real
understanding of software startups on analytics, including instrumentation of analytics, experimentation, diagnostic analysis, and
getting insights. The first concept describes how startups set up
analytics and the latter three illustrate the usage scenarios of analytics.
This study is the first step toward understanding analytics in the
software startup context. The identified concepts can guide further
investigation of analytics in this context. It also provides some insights for software startups to set up analytics for data-informed
decisions. Given the limitation of the data used in the study, the
immediate next step is to ground as well as validate the acquired
understanding using the primary data, by directly interacting with
software startups 
\end{abstract}



\maketitle

\section{Introduction}
In recent years, there has been an increasing interest in utilizing analytics to produce value in the software businesses~\cite{mikalef2018big, guerrouj2016software}. Oftentimes, the power of analytics is used to analyze significant information about software projects ~\cite{guerrouj2016software}. This vast amount of information leads to actions and decisions, transforming ways to revisit business methods and practices~\cite{pappas2018big}. Particularly, software businesses avail this information to understand and make decisions while the project evolves ~\cite{guerrouj2016software}. The use of analytics to facilitate the decision-making process has begun to proliferate since its inception~\cite{mikalef2018big}. Mainly, established software businesses and companies are harnessing the full potential of analytics, possibly because of the access to a large amount of information and abundant resources.\par 
Startup companies are distinguished by their focus on innovation, under extreme uncertainty ~\cite{blank2020four,ries2011lean}. These companies constantly look for a sustainable and scalable business model with speed and focus. Software startups are types of startups, aiming to build software-intensive products or services~\cite{unterkalmsteiner2016software}. Similar to general startup companies, while aiming at innovation, software startups also need to move and act fast. Nevertheless, they are confronted with a few other challenges, related to the software engineering field. Software startups use lean startup methodology to validate the product-market fit ~\cite{bosch2013early} and agile development methodologies to develop the product ~\cite{pantiuchina2017software}. This product and business development cycle certainly brings a change in direction. It is referred to as pivoting in software startups ~\cite{ries2011lean} and brings a change in the product as well as the business model.  We, therefore, may obtain a pertinent illustration that startups adjust themselves based on the information they gain. The decisions regarding directions could be diverse, like for example, changing the target customer segments, adding or removing product features, improving user experience, modifying customer acquisition strategies, and so forth. This is where software startups could utilize analytics, to better understand and evaluate their actions. \par

The research regarding software startups, to date, has primarily tended to focus on software development methodologies~\cite{unterkalmsteiner2016software}. There is still uncertainty, however, how software startups understand and apply analytics throughout the product as well as the business development. The role of analytics in startups is unknown despite the widespread use of analytics in other businesses~\cite{rafiq2021analytics,berg2018role}. Therefore, the current study seeks to address this gap. It will eventually help to understand how startups can use the power of analytics to make data-informed decisions and illuminate their ways towards success. The following Research Question (RQ) is guiding our research: \textit{ RQ: How is analytics understood in the software startup context? }\par

The RQ turns our study exploratory. We utilized secondary data, in the form of text documents, from analytics platforms for startups, to answer the RQ. This study contributes to several important areas. It is the first step towards the understanding of analytics by software startups and eventually explains how do they leverage their decision-making using analytics. Moreover, the findings identify important areas of product and business development where analytics can be applied. 

\section{Related Work}
There are relatively few studies on analytics in the software startup context. In recent work, Rafiq et. al~\cite{rafiq2021analytics} studied how software startups deal with the information to make decisions from an analytics perspective. They reported ten types of recurrent analytics mistakes that might derail software startups. The information dealing mistakes are further grouped into four categories: information collection, information analysis, information communication, and information usage.\par

In another related study, while investigating the role of data analytics in startup companies, Berg et al.~\cite{berg2018role} presented challenges and barriers faced by startups. The study claims that startups are aware of the benefits of applying analytics, however, they are also facing challenges in implementing it. The reported challenges are scarcity of resources, lack of training/skills, time management issues, privacy issues in dealing with the data, and lack of access to the data. It is further argued that the amount of data that startups collect, in the early stages, is insufficient to apply analytics. We find that this study analyzed hardware startups only and big data analytics was brought into the question. This is done possibly because the term \textit{big data} is considered as one of the key aspects of business analytics, however, it differs from traditional data in terms of velocity, volume, and variety~\cite{davis2014beyond}.\par

On the other hand, we find some writers reporting on metrics and measurements in startups. One such example is the multiple case study of Kamulegeya et al.~\cite{kamulegeya2018measurements}, where authors report 28 metrics that startups are utilizing or wish to utilize. The study ~\cite{kamulegeya2018measurements} classifies metrics list according to the measurement practices in established software companies and present five categories: business metrics, product metrics, organizational performance metrics, project metrics, and design metrics.\par

Later, Kemell et al.~\cite{kemell2019100+}, in their multi-vocal literature review, report more than 100 metrics that startups can use. However, we observed that most of these metrics are appropriate for the late stages in the life cycle of startups. Moreover, the study does not reflect on the information needed and the process to collect/track these metrics.\par

Taken together, the studies presented, thus far, provide evidence that there is a lack of understanding on how startups understand the term analytics and apply analytics in the startup context. The current study seeks to address this gap. \par

\section{Research Method}
\begin{figure}

    \includegraphics[width=\columnwidth]{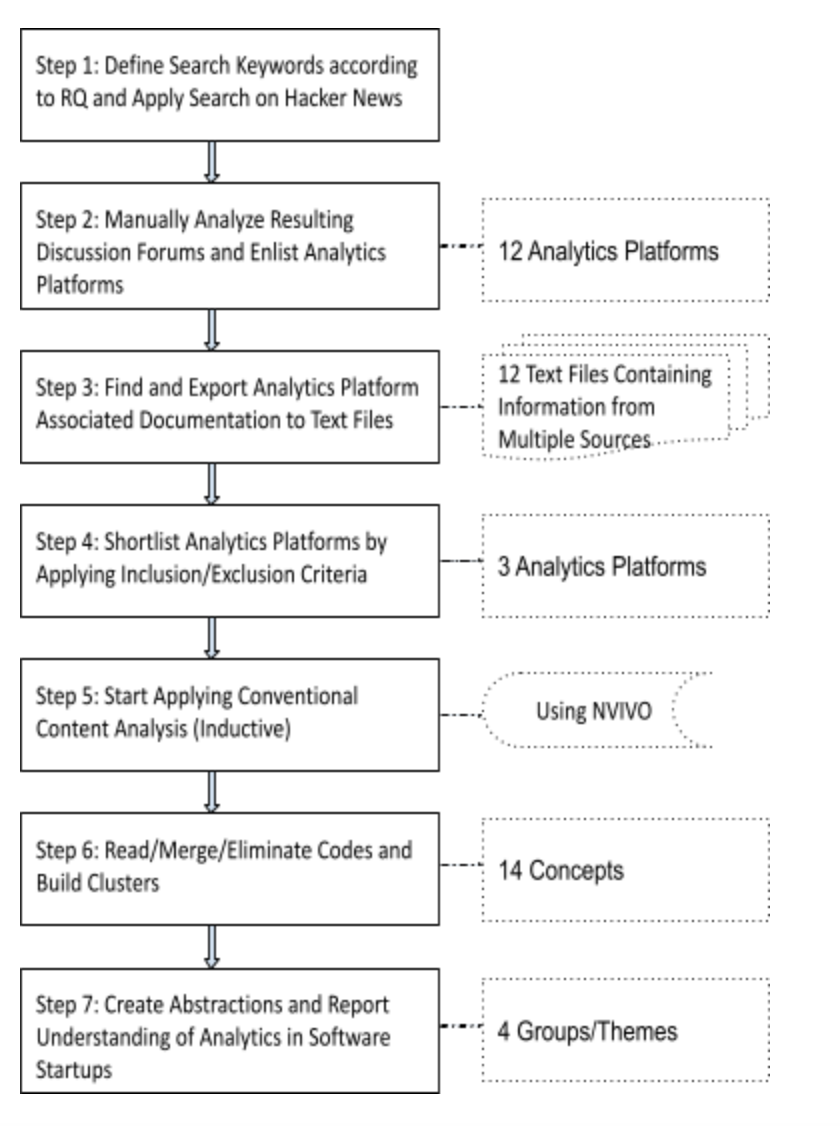}
    \caption{Data Collection and Analysis Procedure}
    \label{fig:procedure}
\end{figure}

Our research is exploratory in nature as the knowledge on the understanding of software startups regarding analytics is limited. Therefore, to develop an initial understanding of what software startups think about analytics, we decided to utilize secondary data i.e. text data from analytics platforms. Nowadays, the widespread usage of secondary data in research makes it a feasible approach to meet limited-time and resources~\cite{johnston2017secondary}.
Vartanian~\cite{vartanian2010secondary} describes that secondary data is the data collected by someone other than the researcher with some other objective. Likewise, to analyze the data set, we used content analysis, a qualitative research method~\cite{hsieh2005three}. A major advantage of using content analysis is that it is considered an appropriate method in the absence of existing theory or when the knowledge on the phenomenon is limited. In particular, we employed a conventional content analysis approach to build an understanding of  analytics in the startup context.  Fig.~\ref{fig:procedure} depicts the complete process of data collection and analysis. The following subsections describe how we collected, filtered, and analyzed the data. 

\subsection{Data Collection}
Based on the research question, we planned to study analytics platforms that are particularly intended for software startups. As a preliminary step on getting information for platforms that startups are using., we consulted forums on Hacker News \footnote{https://news.ycombinator.com/}. Hacker News is the most trusted, reliable, and prevalent source of news for the entrepreneurship and computer science community. It facilitates finding the latest information on startups which is hard to find elsewhere.\par

We applied search terms like "analytics", "metrics" and "startups" to drive the search on the forum. These terms are originated from the research question of this research. Consequently, the manual analysis of the resulting discussions on forums revealed several analytics platforms. These platforms were brought up into a discussion of the startup community, illustrating its features, usage scenarios, and a short comparison. These platforms were ranged from open source to proprietary solutions and covered the following: 
\begin{itemize}
 \item Simple Analytics (\url{https://simpleanalytics.com/})
  \item Mixpanel (\url{https://mixpanel.com/})
 \item Google Analytics (\url{https://analytics.google.com/})
 \item Matomo(\url{https://matomo.org/})
 \item Fathom (\url{https://usefathom.com/})
 \item Plausible (\url{https://plausible.io})
 \item Open Web Analytics (\url{https://www.openwebanalytics.com/})
 \item Kissmetrics (\url{https://www.kissmetrics.io/})
 \item Amplitude (\url{https://amplitude.com/})
 \item Umami (\url{https://umami.is/})
 \item Goatcounter (\url{https://www.goatcounter.com/})
 \item Snowplow (\url{https://snowplowanalytics.com/})
\end{itemize}
Thereafter, we visited individual platform websites, manually extracted the information, and stored it in text files. The information stemmed from several sources, like for example, associated documentation, configuration information, developer information in case of an open-source tool,  user interface, testimonials of customers, and blog posts. We also found blogs of some of the platforms where the vendors expressed their opinions about analytics and shared their experiences in instrumenting analytics at other startup companies. We read each of the available documentation and assessed whether the data can be utilized to address the research question. It is noteworthy to state that we utilized every possible way to extract the fragmented documentation associated with these platforms with an intent not to miss some useful information. To achieve this, we searched platform-specific documentation on the platform website, Hacker News 
 forum and Github \footnote{https://github.com/}, a repository hosting service of open source applications. In this regard, we used the platform's hosting documentation on Github, known as a wiki to extract some more information.\par
 
To obtain another triangulation in the data collection and sampling phase, we explored the website of G2 \footnote{https://www.g2.com/}. G2 is a legit source to check the reviews of business software by real users. We found some of the platforms available on G2, however, few platforms were not listed on it. This has helped us in screening analytics platforms for the final data set. Lastly, we decided to utilize the information from the following three platforms for our data set:
\begin{enumerate}
 \item  \label{itm:amp}Amplitude (\url{https://amplitude.com/})
  \item \label{itm:mix}Mixpanel (\url{https://mixpanel.com/})
 \item \label{itm:pla}Plausible (\url{https://plausible.io})
\end{enumerate}
We considered the following inclusion/exclusion criteria while screening the platform information for our final data set: (1) Platform is mature i.e. used by a huge startup community. We subjectively assessed these criteria from Hacker News 
forum and objectively from stargazer on Github website 
(2) Platform is available on G2 website
, (3) Platform is having extensive documentation, from multiple sources like, for instance, website, blog posts and information from forums. (4) Platform's vendor is/was a software startup. We checked startup information from Crunchbase \footnote{https://www.crunchbase.com/discover/organization.companies}. (5) Platform's documentation contained data wherein at least one or multiple stories were shared. These stories communicate experiences of  software startups while implementing analytics. We assessed this information from the collected documentation of the platform and later on checked individual startups on Crunchbase. In total, the data contained 9 startup stories focusing on analytics, with varying text lengths.  


\subsection{Data Analysis}
We used the qualitative analysis method, named, content analysis to analyze the data. The use of content analysis in the software engineering context is not new to software engineering research. Recent studies, such as~\cite{defranco2017content} and~\cite{kurtanovic2018user}, advocate a rigorous process for its use in software engineering.\par  
Content analysis can be applied to the text data other than interview data and coding categories are directly derived from the data~\cite{hsieh2005three}. This is what makes it different from thematic analysis.  Therefore, in the content analysis research method, text data can be obtained from interviews, focus groups, narrative responses, survey questions, books, articles, or even manuals~\cite{hsieh2005three}. 
In particular, we applied conventional content analysis inductively. This approach is recommended when there is a lack of research knowledge on the topic~\cite{hsieh2005three}.  While performing analysis, we read the whole text repeatedly to get a complete sense of the data. We then started coding phrases wherein the term analytics was indicated or the impression of analytics was arising. 
We highlighted the text that appeared to contain the relevant terms and described the concepts. Subsequently, we utilized the author's words to code this segment of the text. Alongside, we ignored the summative indication of analytics in the text whereby the text was mainly referring to the names of analytics platforms. Meanwhile, we avoided creating new codes as the analysis process continued. It further depicts that we only added new codes when we assessed that the existing codes are not fit to label the new text under consideration.\par

Likewise, soon after finishing the development of codes, we then moved the focus of our analysis to the coded data and examined each code. We eventually examined the text within the particular code and then merged, renamed, and eliminated codes while the analysis process iterated. Based on the similarity in the codes and the relationship between the codes, we created the clusters, eventually organizing clusters into a hierarchical structure. As a result, we obtained 16 themes that were further clustered into 4 high-level themes/top-level categories.\par

We used NVIVO \footnote{https://www.qsrinternational.com/nvivo-qualitative-data-analysis-software/home/}, a qualitative data analysis tool, to support the conventional content analysis. 
 
\section{Findings}
\begin{figure*}
    \centering
    \includegraphics[width=0.8\textwidth] {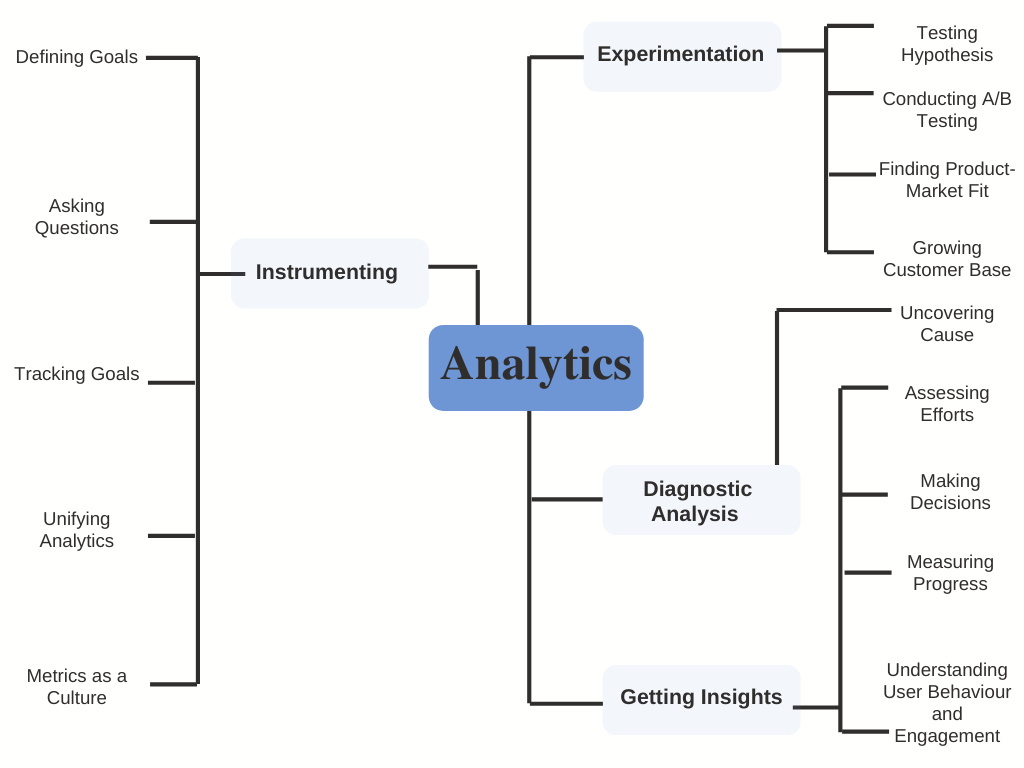}
    \caption{How Software Startups Understand Analytics }
    \label{fig:findings}
\end{figure*}

In this section, we present our research results. We identified 14 themes, in our data set, that directly express or found more relevant to the analytics process inside software startups. We categorized these themes into 4 high-level themes. The high-level themes include instrumenting, experimentation, diagnostic analysis, and getting insights. The clustering of themes and labeling into high-level is purely inductive and thus based on two factors. The first one focuses on how to set up analytics in startups and we name it instrumenting analytics. What we mean by instrumentation of analytics is the intuition and directions to set up analytics in startups. On the flip side, three themes coexist together and focus what are the possible scenarios of applying analytics in startups. Here, we include experimentation, diagnostic analysis and finally getting insights. These themes, when taken together, illustrate how the term analytics is understood by software startups.      
Fig.~\ref{fig:findings} shows how software startups handle analytics. 
\subsection{Instrumenting Analytics:}
\subsubsection{Defining Goals:}
We figured out that all of the startups, employing analytics, have the sense of defining one or multiple goals to achieve, through analytics. Often analytics platform vendors also found promoting culture to establish the goals to meet the business objectives. As an example, in the report of Amplitude platform (\ref{itm:amp}), the concept is explained like this: ``\textit{ Too often, companies develop feature after feature without thinking about how those features meet overall business objectives.}``.  Therefore, these goals might be tied directly to the success of the startup. This further illustrates the thinking of measuring what matters the most and avoiding getting indulged in the noise. It is evident from one of the following excerpts, of Amplitude Platform (\ref{itm:amp}):  ``\textit{Measuring anything and everything leads to unmanageable data. It shifts the burden to your team to try and make sense of the results.}``. In another instance, the company advised in a similar context:``\textit{keep your end goals in mind}``.  \par

Going in the same vein, one of the startups, wanted their clients to hit on the platform. The goal was to increase customer retention. The possible ways they found against the goal were to reduce churn rate and increase customer conversion. Churn rate is the rate at which customers discontinue using a service or product and conversion means how new customers take a desired action~\cite{croll2013lean}. In the data of Mixpanel Platform (\ref{itm:mix}), a startup described the situation in the following words:``\textit{For each market, they have targets they want partners to hit, so they know they are reducing churn and increasing conversion}``. A similar conclusion is echoed in another report of Plausible Analytics Platform (\ref{itm:pla}), where the company expressed its thoughts about setting goals: ``\textit{Most web analytics allow site owners to set goals and events to track those visitor actions that matter the most to them.}``. It is worthwhile to report that these established goals generate several metrics to monitor. The startup reflected this opinion in the following excerpt: ``\textit{goals and events can be tied directly to the monetary success of an enterprise which makes them essential metrics to follow to understand the state of the business}``.

\subsubsection{Asking Questions:}
The data analysis, on the whole, suggests that asking questions is a good way to establish the key focus of analytics, revealing what one wants to achieve through it. And, certainly, this has a connection with setting goals while instrumenting analytics in the startup. In the report of Mixpanel Platform (\ref{itm:mix}), one startup emphasized it in the following way: ``\textit{We need to be asking these questions}``. The same report embarked again: ``\textit{We want people to engage with data, ask questions, and find the answers in data to make the right decisions}``. Likewise, Amplitude Platform (\ref{itm:amp}) data, reported a similar conclusion in the following excerpt: ``\textit{Businesses often have a multitude of questions about their customers and how their product is performing}``.\par

Based on the data set, we encountered the following set of questions that can be asked while instrumenting analytics:What exactly works and what does not?, What do we want to be measuring?, What’s the useful metric?, Is there an event we should be publishing?, Which features are popular?, Which users retain best?, which types of users stick around, and for how long
Who are my most valuable customers?, What actions do those valuable customers take?, Who are the customers who churn?, How many power users you have, What actions do churn users take?, What are the characteristics of highly engaged users?, What are my users’ pain points?, What's happening on your site live in that moment, What are people doing when they visit the website? and What’s happening and how things are going?

\subsubsection{Metrics As a Culture:}
We found several interesting examples in the data where one startup repeatedly emphasized embedding metrics in the overall culture of the product development process. It means that no product feature should be left without publishing different metrics. It is aimed at serving the purpose of measuring and working towards large business goals. The same startup, said on one occasion, in the report of Mixpanel Platform (\ref{itm:mix}): ``\textit{Publishing the right business metrics must be integrated into your development process}``. The speaker, continued: ``\textit{Baking this into the culture and the expectation is probably the most important and most impactful thing that you can do}``. He further advised startups: ``\textit{build it into the culture and build it into the development process so that it becomes a given}``. While establishing this culture, it is necessary that all product platforms i.e. web and mobile, should be aligned and the same vocabulary should be put in place across the development life cycle. This alludes to the following notion: ``\textit{Ensure that all your stakeholders–both in web and mobile–are aligned on how to name those events and properties. }``. A similar conclusion is highlighted in the report of Amplitude Platform (\ref{itm:amp}), where the company said: ``\textit{They align on target product outcomes, define an event taxonomy to measure those outcomes, and instrument tracking code}``. 

\subsubsection{Tracking Goals:}
We observed that another theme in the data is usually concerned with tracking the goals. 
This theme is concerned with tracking established goals. For example, a startup set up a goal and wanted partners to grow. In contrast to this goal, the startup asked many questions and then tracked metrics. This is shown in the following excerpt of a startup, in Mixpanel Platform (\ref{itm:mix}) report: ``\textit{... track how well their restaurant partners are doing on the platform}``. ``\textit{and being able to look at these metrics is what’s going to enable us to do that}``, the startup reported. Another startup, in the same report, had a similar goal and wanted to know the most performing customer acquisition channels and increase customer conversion. The startup reflects its tracking of this goal in the following words:  ``\textit{...been able to measure and optimize high-performing acquisition channels, and also improve purchase funnels, making it easy for consumers to convert into loyal policyholders}``. A similar indication about tracking goals about customer conversion is found in the report of  Plausible Analytics Platform (\ref{itm:pla}), where the startup indicated to track several metrics against goals: ``\textit{Track events and goals to identify the number of unique converted visitors, the total number of conversions, the conversion rate, and the referral sites that send the traffic that converts the best.}``.
\subsubsection{Unifying Analytics:}
Unifying Analytics is one of the challenges that startups must tackle while instrumenting analytics. Based on the findings, we highlight two types of unification, a startup can achieve. The first type incorporates unifying analytics on all of the fragmented products of the startup. Oftentimes, a startup offers multiple platforms, such as a combination of web and mobile app. This is echoed by one of the startups, in the report of Mixpanel Platform (\ref{itm:mix}): ``\textit{the company needed to unify its fragmented analytics ecosystem}``. 
\par
On the flip side, the other aspect of unification discussed in the data is the shared understanding and goal alignment among different startup teams, like, for instance, between the engineering team and product team. The shared understanding and alignment might include target product outcomes, taxonomy to measure outcomes, and instrumenting coding accordingly. The report of Amplitude Platform (\ref{itm:amp}) shows: ``\textit{Implementing great product analytics requires product and engineering teams to work together. They align on target product outcomes, define an event taxonomy to measure those outcomes, and instrument tracking code}``. This enables the engineering team to track only what is required by the product team. The company, Mixpanel Platform (\ref{itm:mix}), agrees with such alignment in the following words: ``\textit{Ensure that all your stakeholders–both in web and mobile–are aligned on how to name those events and properties. Finally, make certain that everyone is bought in on maintaining best practices}``. 

\subsection{Experimentation}
\subsubsection{Testing Hypotheses:}
A recurrent theme in the data was the utilization of analytics in testing hypotheses. Startups need to quickly test several hypotheses. Various startup development approaches like, for example, lean startup and customer development methodology are amplifying the use of Hypotheses, particularly at early stages~\cite{ries2011lean,bosch2013early}. We found startups using analytics to test the hypotheses. Simultaneously, analytics platforms are providing supporting features to achieve this goal. For example, the Amplitude Platform (\ref{itm:amp}) report indicated such features.  Identifying the Winning features,  customer acquisition campaigns, referral sources, and hypotheses regarding customer behavior are prominent in the data. At one instance, the report of Amplitude Platform (\ref{itm:amp}), indicated:``\textit{Its growth engine generates hypotheses data by observing customer behavior while also amplifying winning features and campaign ideas}``. We encountered many instances in the data where this feature was highlighted. It shows that the concerns regarding hypotheses testing through analytics are widespread among startups.  This is evident from the following excerpt of a startup in the report of Mixpanel Platform (\ref{itm:mix}): ``\textit{We wanted to test this hypothesis, so we quickly threw events}``. The startup continued alluding about the outcome of this testing: ``\textit{ we were able to test and disprove our hypothesis of it being related to payday”}``. The testing of the hypotheses saved the startup from investing more effort and time. The startup commented:``\textit{This finding was helpful... because it showed them that, since restaurants were already using the feature, perhaps they didn’t need to invest more in it}``.

\subsubsection{Conducting A/B Testing:}
In the data, at various places, startups expressed the use of analytics in conducting A/B testing. They illustrated the benefits, this testing has brought to them and saved them in several ways. We found two examples where startups measured and followed the results of A/B testing through the use of analytics. Mainly, A/B tests were conducted to test the focus of the product's features. At the same time, one startup instrumented the process of analytics in their environment because they needed to clarify and learn from A/B tests. The startup reports this finding in the report of Mixpanel Platform (\ref{itm:mix}) using the following words: ``\textit{The company required a tool that could quickly clarify and streamline the learnings from A/B testing}``. The same startup continued expressing its interests in A/B testing measurements through analytics and said that this helped them to improve account creation using A/B testing. It is expressed in the following excerpt:``\textit{monitor an A/B test to see if it could increase another important metric: New Signer Accounts}``.\par

While, on the other hand, in the Amplitude Platform (\ref{itm:amp}) report, a startup argued that analytics helps in A/B testing regardless of the exclamation points the test generates. The startup commented: ``\textit{Analytics are for more complex than A/B testing whether two exclamation points or three generate more opens}``. The startup further commented: ``\textit{Analytics helped ... his team shift their entire messaging strategy toward what their users valued}``.

\subsubsection{Finding Product-Market Fit:}
A common view expressed in all the reports is the use of analytics to truly understand the target customers and move towards the development of the right product for them. This means that startups can assess analytically whether their business model is flawed or has the potential to grow fast. In several instances, in the data of Amplitude Platform (\ref{itm:amp}), we encountered the company standing out in explaining the similar benefits. For instance, at one place, the company reported:``\textit{analytics is the epicenter of how digital-first companies figure out customer needs and measure the impact of their products}``. It continued expressing significant use of analytics in the product-market fit for startups: ``\textit{ To improve your odds of finding product-market fit, you need an analytics platform that can go past reporting vanity metrics and basic conversion funnels to understanding the experience customers are having in your product}``. In the same way, another company, Plausible Analytics (\ref{itm:pla}) also reported building an understanding of the product benefits. The company alluded to the view in the following words: ``\textit{Understanding the interest around your product’s benefits..}``. 

\subsubsection{Growing Customer Base:}
We find in our analysis that, a growing customer base is regarded as the experiment-driven process, whereby different options are built and tested. We report an interesting event when a startup aimed to remove pop-ups from its application to increase the customer activation rates. After three months of onboarding, they accidentally removed the pop-ups even when they were not ready to embrace this change. The accident went unnoticed for two weeks by the startup and suddenly they realized a massive increase in the customer activation rates. A deep dig down through the analytics platform revealed pop-up removal as the main reason for this. In the report of Amplitude Platform (\ref{itm:amp}), the startup explained this event in the following words: ``\textit{We talk about growth and retention as “experiment-driven” processes so much that we sometimes lose sight of the fact that many discoveries occur by chance}``. The startup regarded customer growth as an experiment-driven process, as they also planned to test it in the same way. Now, in a retrospect, they sense that although it was an accident, however, without the analytics the root cause cannot be exactly known. The startup explains this in the following excerpt: ``\textit{They can only happen, however, if you’ve been tracking the data you need to work back to the root cause later}``. They continued sharing their pleasant accident:``\textit{Without data to look back on, this whole affair would have been regarded as a massive error }``. 
\subsection{Diagnostic Analysis}
\subsubsection{Uncovering Cause:}
The data analysis revealed that software startups are utilizing analytics to diagnose a strange or unexpected cause of a phenomenon. The strange situations may include sudden spikes in application traffic at particular times or a sudden decrease in customer conversion rate. Customer conversion rate is the percentage of users who take the desired action on the product\cite{croll2013lean}. We found 3 instances in the reports of two analytics platforms, where software startups reported their experience in finding the cause of a strange phenomenon just because they were already into the analytics. For instance, we found that one of the startup customers of Amplitude Platform (\ref{itm:amp}) , noticed an unusual yet recurring trend of a spike in traffic. They were new to applying analytics but they managed to sort out the reason eventually. And they assessed that it is the start of the weekend and the traffic is spiking every Monday or on Tuesday. The startup figured out, with the help of analytics, that their partners were looking at their feedback, sales of last week, changes to opening hours or menus, at the beginning of every week. 
The startup explained the diagnosis:``\textit{The company continued to collect a month’s worth of data and realized that DAU was spiking every Monday—or, in the case of a long weekend—Tuesday. It’s the start of the work week and of course our partners are logging in}``. Here, DAU is the abbreviation for daily active users.\par

A similar situation is reported by another startup at Mixpanel platform (\ref{itm:mix}). The startup experienced a sudden drop in the conversion rate and later it was found that a minor change on the home page has caused this drop. The startup also utilized analytics in finding the cause. According to the Mixpanel (\ref{itm:mix}), ``\textit{If you weren’t paying close attention to your analytics, you would just wind up sitting around and wondering why your conversion rate had dropped}``. 

\subsection{Getting Insights:}
\subsubsection{Assessing Efforts:}

In this theme, startups obtained insights on the efforts that they put in terms of introducing new product features, making changes in the product layout, and hiring variant customer acquisition as well as marketing channels. They want to observe which of their efforts are more result-oriented so that they can shift all their focus to the identified direction. This new direction leads the future strategy of the startup and brings better understanding the customer needs. As, one startup, Plausible Analytics (\ref{itm:pla}) indicated:``\textit{You may be putting a lot of time, effort, and resources into different marketing campaigns and by looking at referral sources of your website traffic you can better understand which of those campaigns are more worth than others}``. The startup continued arguing that one needs to redesign its future strategy if the existing efforts are misleading. ``\textit{If you’re spending a lot of time and effort on a community but that effort doesn’t result in any benefits to your site or business, then you need to reconsider things}``, the startup remarked on the situation. While talking about the future strategy, the startup asserted to analyze efforts to optimize the future. It is reported in the following phrase: ``\textit{an addition layer and more depth to analyze your efforts which can then help you optimize your strategy for the future}``.\par

A similar experience is shared in the report of Amplitude Platform (\ref{itm:amp}), where the company highlighted the need for analytics to deeply understand the customer needs and assess the outcomes of development efforts. The statement reflects this finding: ``\textit{The one thing we found missing was a powerful product platform to truly understand what users wanted and the impact our development efforts were having on their user experience}``. 

\subsubsection{Measuring Progress:}
Savvy startups are often concerned with measuring their progress. However, measuring the progress without any yardstick is barely possible. This is what we pointed out while analyzing the data. We found that startups are using analytics to measure their progress using insights from different perspectives. The startup, Plausible Analytics (\ref{itm:pla}), told in their report: ``\textit{we use web analytics to measure our startup's progress and make better decisions}``. Generalizing this need, the startup claimed: ``\textit{majority of website and business owners want to see some level of stats that tells them what’s going on }``. The company further highlighted the key advantages of using analytics in measuring progress, in following excerpt: ``\textit{some of the main areas where web analytics can help website owners get a better idea of what’s happening and how things are going }``.

\subsubsection{Making Decisions:}
The data analysis depicts that, startup decisions are based on the insights that the analytics generates for them. These insights oftentimes arise by measuring the progress of startup, by assessing the startup efforts in making things done, or during the experimentation, however, these are considered holistically while making decisions. For instance, Plausible Analytics (\ref{itm:pla}) claimed that they used analytics in making decisions. These decisions further lay down the foundation of future strategy. It is apparent from the following excerpt: ``\textit{... more depth to analyze your efforts which can then help you optimize your strategy for the future }``. In the same way, a startup, in the report of Mixpanel Platform (\ref{itm:mix}), shared that they decided not to put more effort into designing features when they observed that their partners are already engaged with the product. The startup commented: ``\textit{This finding was helpful ... because it showed them that, since restaurants were already using the feature, perhaps they didn’t need to invest more in it.}``. The company Amplitude Platform (\ref{itm:amp}) sheds more light on making decisions employing analytics in these words: ``\textit{Your product analytics can provide you with the exact data needed to drive your decision-making and continue to make improvements that delight your customers. }``.

\subsubsection{Understanding User Behaviour and Engagement:}
One possible use of analytics, our data analysis showed, is understanding user behavior and interaction with the product. It is used to clear the assumptions regarding the customer's behavior. At the same time, it brings to the surface, trending content, winning product features, and highlights the customer journey from one platform to another. Several analytics goals can be achieved through this, like, for example, improving retention by monitoring and understanding user behavior. The report of Amplitude Platform (\ref{itm:amp}), highlights this finding in the following words: ``\textit{Identifying which customers are most engaged and using this information to improve retention }``. The company continued expressing similar and concrete benefits at another place in the following excerpt: ``\textit{analytics helps you create that digital experience without any guesswork. It provides you with concrete information to optimize conversions, grow retention, and maximize revenue. }``. \par

Another interesting finding, we noticed, is finding reasons of a user behaviour. ``\textit{Good product teams don’t just ask, what are my users doing? They ask, why? This requires context}``, the report of Amplitude Platform (\ref{itm:amp}), revealed. 

\section{Discussion}
Our study results confirm that analytics in the startup context reflects a somewhat different meaning in contrast to what is described in the software engineering literature. Therefore, when we talk about startups, analytics is not based on the big data as it is indicated in ~\cite{berg2018role,mikalef2018big}, nor it is characterized by what is given in terms of software analytics ~\cite{guerrouj2016software}. However, we find that while there exist many types of analytics in the literature, e.g. big data analytics, web analytics, software analytics, and social media analytics, startup analytics is closer to the characteristics of web analytics. Alongside, we need further investigation to ground the understanding of analytics in the startup context. \par

The study~\cite{rafiq2021analytics} also reported a few other analytics-related mistakes, e.g.  ``\textit{avoiding to collect further information}``, ``\textit{poor team communication}`` and ``\textit{mishandling information}``. Our data shows that the first two mistakes can be mitigated by instrumenting analytics in startups. For instance, unifying analytics will reduce poor communication in the startup team. Similarly, for the last one, the findings suggest looking at information from multiple perspectives and also waiting for trends to appear in the information. \par

Surprisingly, one of our findings emphasizes developing a culture of metrics and embedding it into the development of product features. This invites the concept of measuring everything, especially in product development. A possible explanation for this result might be that startups could need to produce hindsight. Therefore, in a retrospect, they might need to do this. However, we find no clue on this finding from literature, and our study also remains unable to explain it to a further extent.  \par    

In contrast to the earlier findings of~\cite{berg2018role}, however, we confirm that software startups are using analytics. This difference may be explained by our chosen sample. We studied purely software startups while Berg et al. ~\cite{berg2018role} studied startups with a hardware part as well. Another important difference, we would like to illustrate is that our study focused on the understanding of startups about analytics. We find that, when taken together, this understanding is pretty different from what general software engineering literature states about analytics. The literature~\cite{mikalef2018big, guerrouj2016software,pappas2018big}, presents analytics in terms of big data analytics, software analytics, or simply data analytics. It is interesting to reveal that our findings provide a diverse understanding of analytics from a startup's perspective. However, it would be interesting to relate our findings with general software engineering literature. This would remain an important issue for future avenues. \par 
Lastly, the findings strongly show a connection of analytics with metrics. For example, goals and questions during analytics instrumentation generate metrics. However, our findings are not self-explanatory in explaining this connection and further assessment of the relationship between analytics and metrics. 

\subsection{Threats to Validity}
One of the threats to validity, particularly, internal validity, is the use of secondary data. This type of data provides a lack of control on data volume and quality. To mitigate this threat, to a certain extent, we triangulated data using multiple sources and by applying inclusion/exclusion criteria. This type of data has already been used by other studies, like for example, by ~\cite{auer2020continuous}.\par

Regarding threats to external validity, one of the threats lies in our data collection strategy. We selected 3 platforms out of 12 for further investigation by applying several inclusion/exclusion criteria. Later, we collected analytics platform documentation from every possible source e.g. platform website, blogs, and forums. The data, from the three platforms, was different in length and scope. Besides that, we might have missed collecting some associated data. However, the data set was containing startups from different geographical regions, market segments, and different product platforms. It enhanced the ability to generalize the results of this ongoing research effort.  \par

Regarding the reliability of this study, one possible threat is concerned with the researcher's bias in the coding process. The coding process was done by one researcher solely, however, the early codes and themes were discussed with the other researcher before reporting.\par

Lastly, we take the privilege to admit that the reported understanding of analytics, in the startup context, still needs further investigation, possibly with the primary data. It further means that we need to continue grounding and validating our understanding of analytics in the startup context based on the primary data.  

\section{Conclusions and Future Work}
Startups are confronted with several challenges while raising the odds of success. Uncertainty, scarcity of resources, engineering challenges, speed, and right focus are among few to mention. On their way, startups have to take plenty of decisions and adjust directions accordingly. This is where analytics can serve them with the right information. This is the first study that attempts to explain how analytics is understood in the startup context. In this ongoing research, we report an initial understanding of analytics in the startups covering its instrumentation and the context, in which it can be utilized. \par

Several questions still need to be answered in the future. One significant direction is to validate this understanding of analytics in startups with the primary data. Moreover, the current findings have left many questions in need of further investigation. For example, further work is required to map the relationship of \textit{analytics} with the term \textit{metrics}. As both these terms are used interchangeably. Lastly, the relationship between stages of startup across the life-cycle and use of analytics might be worth investigating.

\bibliographystyle{ACM-Reference-Format}
\bibliography{main}
\end{document}